\documentclass[twocolumn,showpacs,preprintnumbers,amsmath,amssymb]{revtex4}
  \usepackage[latin1]{inputenc}
  \usepackage{amssymb}

  \usepackage{graphicx}

  \graphicspath{{caract_figures/}}

  \usepackage{color}		% Need the color package
  \usepackage{epsfig}
  \usepackage{ifthen}
  \usepackage{fancyhdr}

  \newcommand{\adraft}{false}

  \ifthenelse{\equal{\adraft}{true}}{
    \newcommand{\rem}[1]{{{\textbf{\textsf{#1}}}}}
    \newcommand{\comment}[1]{\small\textsf{\textbf{(#1)}}\normalsize}
  }{
    \newcommand{\comment}[1]{}
    \newcommand{\rem}[1]{}
  }

  \newlength{\mybasewidth}
\begin{document}

  \title[Quasi-static double-reed characteristics]{Quasi-static non-linear characteristics of double-reed instruments}

  \author{André Almeida}
  \affiliation{IRCAM -- Centre Georges Pompidou -- CNRS UMR9912, 1 Place Igor Stravinsky; 75004 Paris; France.}
  \email{Andre.Almeida@ircam.fr}

  \author{Christophe Vergez}
  \affiliation{Laboratoire de Mécanique et Acoustique -- CNRS UPR7051,
    31 Ch. Joseph Aiguier; 13402 Marseille Cedex 20; France.}
  \email{vergez@lma.cnrs-mrs.fr}

  \author{René Caussé}
  \affiliation{IRCAM -- Centre Georges Pompidou  --- CNRS UMR9912, 1 Place Igor Stravinsky; 75004 Paris; France.}
  \email{Rene.Causse@ircam.fr}

  \date{\today}
  

  %Use showkeys class option if keyword

  \begin{abstract}
    This article proposes a characterisation of the double-reed in
    quasi-static regimes. The non-linear relation between the pressure
    drop $\Delta p$ in the double-reed and the volume flow crossing it
    $q$ is measured for slow variations of these variables. The volume
    flow is determined from the pressure drop in a diaphragm replacing
    the instrument's bore.
%, a technique used in the past for similar measurements in clarinet mouthpieces \cite{Dal03}
    Measurements are compared to other experimental results on reed
    instrument exciters and to physical models, revealing that
    clarinet, oboe and bassoon quasi-static behavior relies on similar
    working principles. Differences in the experimental results are
    interpreted in terms of pressure recovery due to the conical
    diffuser role of the downstream part of double-reed mouthpieces
    (the staple).
  \end{abstract}
  
  \maketitle

\section{Introduction}

\subsection{Context}

% \rem{
%   \begin{itemize}
%   \item Exciter / resonator:
%     \begin{itemize}
%     \item linked through coupling variables
%     \item non-linear relation between coupling variables
%     \item characterises the reed globally in a synthetic manner
%     \item define what's understood by ``the nonlinear
%       characteristics''
%     \end{itemize}
%   \item Our aim:
%     \begin{itemize}
%     \item Measure the non-linear relation (characteristics) in a
%       situation for which the dependency between $p$/$q$ can be
%       considered instantaneous (quasi-static)
%     \item Propose some tentative explanations for the observations
%     \end{itemize}
%   \end{itemize}
% }

% The usual method for studying and simulating the behavior of
% self-sustained instruments is to identify two main systems, a passive
% part (resonator), responsible by imposing its modes of vibration, and
% an active part (exciter), which creates and maintains the oscillations
% in the resonator. 

The usual method for studying and simulating the behavior of
self-sustained instruments is to separate them in two functional parts
that interact through a set of linked variables: the resonator,
typically described by linear acoustics, and the exciter, a non-linear
element indispensable to create and maintain the auto-oscillations of
the instrument \cite{Hel77}.

Although this separation may be artificial because there may not be a
well-defined boundary between the two systems, it is usually a
simplified view that allows to describe the basic functioning
principles of the instrument. In reed instruments, for instance, the
resonator is assimilated to an air column inside the bore, and the
exciter to the reed, which acts as a valve.

In the resonator of wind instruments, the relation between the
acoustic variables, pressure ($p$) and volume flow ($q$) can be
described by a linear approximation to the acoustic propagation which
has no perceptive consequences in sound simulations \cite{Gil05}.  The
oscillation arises from the coupling between the reed and the air
column, which is mathematically established through the variables $p$
and $q$.

% In most reed instruments, the resonator can be described by linear
% relations between two oscillating acoustic variables, pressure ($p$)
% and volume flow ($q$), without perceptive consequences in sound
% simulations \cite{Gil05}. The oscillation arises from the coupling
% between the reed and the air column, which is mathematically
% established through these acoustic variables $p$ and $q$. 

The exciter is necessarily a nonlinear component, so that the
continuous source of energy supplied by the pressure inside the
musician's mouth can be transformed into an oscillating one
\cite{Hel77} \cite{Fle98}.  The characterisation of the exciter thus
requires the knowledge of the relations between variables $p$ and $q$
at the reed output (the coupling region).

In principle this relation is non-instantaneous, because of inertial
effects in the reed oscillation and the fluid dynamics.  A first
approach to the characterisation of the exciter is thus to restrict
the study to cases where these delayed dependencies (or, equivalently,
time derivatives in the mathematical description of the exciter) can
be neglected.

% However, the
% typical delays involved in the exciter relations are usually much
% shorter than those present in the resonator. \rem{(REF)}

% The aim of this paper is thus to give describe the mathematical
% relation between the pressure and volume flow at the reed output, in a
% quasi-static case, that is, when the time variations of $p$ and $q$ can be
% neglected.

The aim of this paper is to measure the relation between the pressure
and volume flow at the double reed output in a quasi-static case, that
is, when the time variations of $p$ and $q$ can be neglected, and to
propose a model to explain the measured relation.

\subsection{Elementary reed model}
\label{sec:backuss-model}

In quasi-static conditions, a simple model can be used to describe the
reed behavior \cite{Wil74}. The reed opening area ($S$) is controlled
by the difference of pressure on both sides of the reed: the pressure
inside the reed ($p_r$) and the pressure inside the mouth ($p_m$). In
the simplest model, the relation between pressure and reed opening
area is considered to be linear and related through a stiffness
constant ($k_s$):
\begin{equation}
  \label{eq:spring}
  (\Delta p)_r = p_m - p_r = k_S (S_0 - S)
\end{equation}
In this formula, $S_0$ is the reed opening area at rest, when the
pressure is the same on both sides of the reed. In most instruments
(such as clarinets, oboes or bassoons) the reed is said to be
\emph{blown-closed} (or \emph{inward-striking}) \cite{Hel77}, because
when the mouth pressure ($p_m$) is increased, the reed opening area
decreases.

The role of the reed is to control and modulate the volume flow ($q$)
entering the instrument. The Bernoulli theorem applied between the
mouth and the reed duct determines the velocity of the flow inside the
reed ($u_r$) independently of the reed opening area:
\begin{equation}
  \label{eq:Bern}
  p_m + \frac{1}{2} \rho u_m^2 = p_r + \frac{1}{2} \rho u_r^2 
\end{equation}
In this equation, $\rho$ is the air density. Usually, the flow
velocity $u_m$ is neglected inside the mouth, because of volume flow
conservation: inside the mouth the flow is distributed along a much
wider cross-section than inside the reed duct.

% Although the flow velocity $u_r$ is constant in a Bernoulli model, the
% volume flow $q$ is not, because it is the integrated flow velocity in
% the reed opening area. 

The volume flow ($q$) is the integrated flow velocity ($u_r$) over a
cross-section of the reed duct. For the sake of simplicity, the flow
velocity is considered to be constant over the whole opening area, so
that $q=S u_r$. Using equation (\ref{eq:Bern}), the flow is given by
\begin{equation}
  \label{eq:Flow}
   q = S \sqrt{\frac{2(p_m - p_r)}{\rho}} 
\end{equation}

Combining eq.~(\ref{eq:Flow}) and eq.~(\ref{eq:spring}), it is
possible to find the relation between the variables that establish the
coupling with the resonator ($p_r$ and $q$):
\begin{equation}
  \label{eq:elem-model}
  q = \frac{p_M - (\Delta p)_r}{k_s} \sqrt{\frac{2(\Delta p)_r}{\rho}}
\end{equation}

\begin{figure}[htbp]
  \centering
  \includegraphics[width=.6\columnwidth]{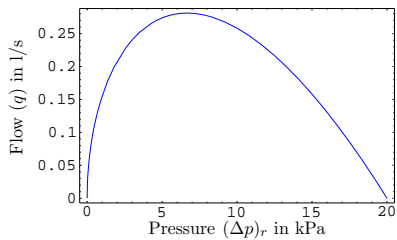}
%  \scalebox{0.8}{\input{caract_figures/caract_sr.pstex_t}}
  \caption{A typical non-linear
    characteristic curve for a reed of dimensions similar to an oboe
    reed, given by equation (\ref{eq:elem-model})}
  \label{fig:car_sr}
\end{figure}

The relation defined by equation (\ref{eq:elem-model}) is plotted in
figure \ref{fig:car_sr}, constituting what we will call the elementary
model for the reed. 

The static reed beating pressure \cite{Dal05} $p_M = k_s S_0$
\footnote{\emph{Static beating pressure}: The minimum pressure for
  which the reed channel is closed} is an alternative parameter to
$S_0$, and can be used as a magnitude for proposing a dimensionless
pressure:
\begin{equation}
  \label{eq:padim}
  \tilde{p} = (\Delta p)_r / p_M
\end{equation}
Similarly a magnitude can be found
for $q$, leading to the definition of the dimensionless volume flow
\begin{equation}
  \label{eq:qadim}
  \tilde{q} = \frac{k_s}{p_M^{3/2}}\sqrt{\frac{\rho}{2}} q
\end{equation}
Equation (\ref{eq:elem-model}) can then be re-written in terms of
these dimensionless quantities:

\begin{equation}
  \label{eq:adim-el-model}
  \tilde{q} = (1-\tilde{p})\tilde{p}^{1/2} 
\end{equation}

This formula shows that the shape of the non-linear characteristic
curve of the elementary model is independent of the reed and blowing
parameters, although the curve is scaled along the pressure $p$ and
volume flow $q$ axis both by the stiffness $k_s$ and the beating
pressure $p_M = k_s S_0$.

\subsection{Generalisation to double-reeds}

% \rem{
%   \begin{itemize}
%   \item simple (generic) models available and suitable for single-reeds \ref{sec:backuss-model}
%   \item characteristic measured for single-reeds
%   \item double-reeds: never measured
%   \item double-reeds: suggested NLC curves in the literature \cite{Hir95} (maybe after the following subsubsection)
%   \end{itemize}
% }

% \rem{Maybe this section should be after the elementary model\ldots
%   this way it would be easier to say that Dalmont's measurements fit
%   to the elementary model}

For reed instruments, the quasi-static non-linear characteristic curve
has been measured in a clarinet mouthpiece \cite{Bac63}, \cite{Dal03},
and the elementary mathematical model described above can explain
remarkably well the obtained curve almost until the reed beating
pressure ($p_M$).

For double-reed instruments it was not verified that the same model
can be applied. In fact, there are some geometrical differences in the
flow path that can considerably change the theoretical relation of
equation (\ref{eq:adim-el-model}). Local minima of the reed duct
cross-section may cause the separation of the flow from the walls and
an additional loss of head of the flow \cite{Wij95}, and in that case
the characteristics curve would change from single-valued to
multi-valued in a limited pressure range. This kind of change could
have significant consequences on the reed oscillations.

However, the non-linear characteristic relation was never measured
before for double-reeds, justifying the work that is presented below.

%  Along the reed duct, there is a
% point where the cross-section reaches a local minimum (which can
% change according to the instrument). 

\section{Principles of measurement and practical issues}
\label{sec:car_difficult}

The characteristic curve requires the synchronised measurement of two
quantities: the pressure drop across the reed $(\Delta p)_r$ and the
induced volume flow $q$.

\subsection{Volume flow measurements}
\label{sec:car_flow_meas}

One of the main difficulties in the measurement of the reed
characteristics lies in the measurement of the volume flow. There are
instruments which can accurately measure the flow velocity in an
isolated point (LDA, hot-wire probes) or in a region of a plane (PIV),
but it can be difficult to calculate the corresponding flow by
integrating the velocity field. In fact it is difficult to do a
sampling of a complete cross-section of the reed because a large
number of points would have to be registered. Supposing that the flow
is axisymmetric at the reed output (which is confirmed by experimental
results \cite{Alm06}), the measurement along a diameter of
the reed would be sufficient, but regions close to the wall are
inaccessible.  

% Another difficulty would be making sure that the
% profile is being measured along a diameter.

% Since the flow is found
% to be approximately axisymmetric at the reed output (see
% sect.~\ref{sec:diamprof_perp}), the measurement along a diameter of
% the reed would be sufficient, but regions close to the wall are
% inaccessible.  Another difficulty would be making sure that the
% profile is being measured along a diameter.

%  line of symmetry of the flow (which may not coincide
% exactly with the line of symmetry of the reed output --- see
% sect.~\ref{sec:daimprof_perp})

% It can be also difficult to make sure that the
% measurement line is in fact a radial line\ldots

% \rem{Some of these problems also exist with the hot-wire}

% but they
% are usually difficult to implement or not sufficiently accurate to be
% able to calculate the flow by integration of the velocity field.

On the other hand, commercial flow-meters usually have the
disadvantage of requiring a direct reading, which would have been
unpractical for a complete characteristic measurement (large number of
readings in a short time interval).

% .
% Since several measurements were required in a relatively short time
% interval, it would be unpractical to use such an experimental
% methodology.

% Commercial flow meters (floating weight, rotameters) have the
% disadvantage of not being sufficiently precise in time to do a
% complete measurement of the characteristic, because in some situations
% conditions can vary faster than the response  time of such instruments.

\begin{figure}[htbp]
  \centering
%   \psfrag{dps}{$\left(\Delta p\right)_s$}
%   \psfrag{dpr}{$\left(\Delta p\right)_r$}
%   \psfrag{dpd}{$\left(\Delta p\right)_d$}
%   \psfrag{p1}{$p_m$}
%   \psfrag{p2}{$p_r$}
%   \psfrag{patm}{$p_{atm}$}
  \includegraphics[width=.6\columnwidth]{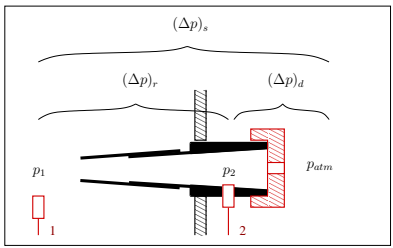}
  \caption
  {Use of a diaphragm to measure flow and pressure difference in the reed. Numbered rectangles correspond to the pressure probes used in the measurement.}
  \label{fig:diaph_sk}
\end{figure}

An indirect way of measuring the flow was then preferred to the above
mentioned methods. It consists in introducing a flow resistance in
series with the reed, for which the pressure can be accurately related
to the flow running through it (see fig.~\ref{fig:diaph_sk}).
% diaphragm -- Reference to S. Ollivier

The diaphragm method, used successfully by S.~Ollivier \cite{Oll02} to
measure the non-linear characteristic of single reeds, is based on
this principle. The resistance is simply a perforated metal disk which
covers the reed output.

% so that the flow is forced to cross the hole in the diaphragm.

% In the particular case of the reed, the diaphragm method has the
% advantage of providing an acoustic resistance that reduces the natural
% tendency of the reed to oscillate (see below).

For such a resistance, and assuming laminar, viscous less flow, the
pressure drop $\left(\Delta p\right)_d = p_r - p_{atm}$ across the
diaphragm can be approximated by the Bernoulli law \footnote{The flow
  velocity at the reed output is neglected when compared to the
  velocity inside the diaphragm ($S_{d} \ll S_{output}$)}:
\begin{equation}
  \label{eq:diaph}
  \left(\Delta p\right)_d = p_r - p_{atm} = \frac{1}{2} \rho \left(\frac{q}{S_{d}}\right)^2
\end{equation}
where $q$ is the flow crossing the diaphragm, $S_{d}$ the cross
section of the hole, and $\rho$ the density of air. In our experiment,
pressure $p_{atm}$ is the pressure downstream of the diaphragm
(usually the atmospheric pressure, because the flow opens directly
into free air). The volume flow $q$ is then determined using a single
pressure measurement $p_r$.

% To calculate the flow, it is thus only necessary to measure the
% pressure behind the diaphragm $p_r$. A second pressure measurement is
% nevertheless needed in order to calculate the pressure drop between
% the mouth and the reed output. A pressure probe is then inserted
% inside the mouth, giving the value of $p_m$. The pressure drop inside
% the reed in then $p_m-p_r$, to be related to the flow $Q$.

\subsection{Practical issues and solutions}
\label{sec:car_pract}

\subsubsection{Issues}
\label{sec:carac_choice_diaph}

The realisation of the characteristic measurement experiments
encountered two main problems:

\paragraph{Diaphragm reduces the range of $(\Delta p)_r$ for which the measurement is possible}
\label{sec:diaphr-reduc-range}

\begin{figure}[htbp]
  \centering
  \includegraphics[width=.6\columnwidth]{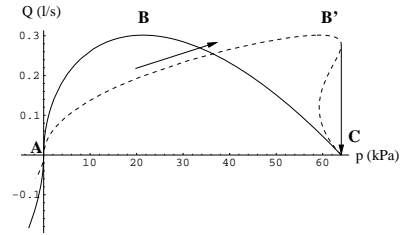}
  \caption
  {Comparison of the theoretical reed characteristics (solid line)
    with the a model of the overall characteristics of the reed
    associated to a diaphragm (dashed) --- mathematical models, based
    on the Bernoulli theorem \cite{Wij95}.}
  \label{fig:sys_car}
\end{figure}

The addition of a resistance to the air flow circuit of the reed
changes the overall nonlinear characteristic of the reed plus
diaphragm system (corresponding to $\left(\Delta p\right)_s$ in
fig.~\ref{fig:diaph_sk} and to the dashed line in
fig.~\ref{fig:sys_car}). The solid line plots the flow against
$\left(\Delta p\right)_r$, the pressure drop needed to plot the
non-linear characteristics. When the resistance is increased, the
maximum value of the system's characteristic is displaced towards
higher pressures \cite{Wij95}, whereas the static beating pressure
($p_M$) value does not change (because when the reed closes there is
no flow and the pressure drop in the diaphragm ($\left(\Delta
  p\right)_d$) is zero).

Therefore, if the diaphragm is too small (i.~e., the resistance is too
high), part of the decreasing region (\textbf{B'C}) of the system's
characteristics becomes vertical, or even multivalued, so that there
is a quick transition between two distant flow values, preventing the
measurement of this part of the characteristic curve \cite{Dal03} (see
figure \ref{fig:sys_car}). A critical diaphragm size ($S_{d,crit}=0.58
S_0$) can be found below which the characteristic curve becomes
multi-valued (see appendix \ref{sec:calc-minim-diaphr}).

% However, since the diaphragm is also used to control the acoustic
% resistance and therefore prevent auto-oscillations, the diaphragm
% should not be too large. Practically, after several tries, no diameter
% was found that would allow the whole curve to be measured while
% preventing the reed oscillations. This issue is addressed in the next
% paragraph.
\paragraph{Reed auto-oscillations}
\label{sec:car_autoosc}

% The restriction of the measurements to quasi-static regimes is a
% simplification in terms of the amount of measured data because it is
% not necessary to measure a different curve for each frequency, and
% because simpler methods can be used to measure the flow in
% quasi-static conditions. Nevertheless, this restriction requires that
% the pressure and flow vary smoothly throughout the whole range of
% measurement.  This can be difficult to achieve in practice, because of
% auto-oscillations, sought in a normal utilisation of the instrument,
% but that have to be preventent with ted here using artificial procedures.

% it is the mechanism that allows the
% production of notes. It occurs for a part of the whole range of
% pressures required for the characteristic measurement. In particular,
% when increasing the pressure, the auto-oscillations will start above
% the threshold of oscillation $P_T$, defined in section
% \ref{sec:dblreed_func} for the Backus reed model. For our present aim
% though, this phenomenon is a drawback, and

% Reason of the auto-oscillations

% 1-The acoustic feedback

Auto-oscillations have to be prevented here to keep consistent with
the quasi-static measurement (slow variations of pressure and flow).
This proved to be difficult to achieve in practice. In fact,
auto-oscillations become possible when the reed ceases to act as a
passive resistance (a positive $\frac{\partial q}{\partial p}$, which
absorbs energy from the standing wave inside the reed channel) to
become an active supply of energy ($\frac{\partial q}{\partial p}<0$).
All real acoustic resonators are slightly resistive (the input
admittance $Y_{in}$ has a positive real part).  This can compensate in
part the negative resistance of the reed in its active region, but
only below a threshold pressure, where the slope of the characteristic
curve is smaller than the real part of $Y_{in}$ for the resonator
\cite{Deb04b}.

One way to avoid auto-oscillations is thus to increase the real part
of $Y_{in}$, that is the acoustic resistance of the resonator. It is
known that a an orifice in an a acoustical duct with a steady flow
works as an anechoic termination works as an acoustic resistance
\cite{Dur01}, so that if the diaphragm used to measure the flow (see
sect.~\ref{sec:car_flow_meas}) is correctly dimensioned, the acoustic
admittance seen by the reed $Y_{in}$ can become sufficiently resistive
to avoid oscillations.

\rem{The resistance caused by the diaphragm ($\mathcal{R}(Y_d)$) could
  be compared to the maximum admittance of the double reed
  ($\mathrm{max}\left(\mathcal{R}(\frac{\partial q}{\partial (\Delta
      p)_r})\right)$) to find appropriate parameters for the diaphragm
}

\subsubsection{Solutions proposed to adress these issues}
\label{sec:car_sol}

\paragraph{Size of the diaphragm}

The volume flow is determined from the pressure drop across the
diaphragm placed downstream of the reed. In practice there's a
tradeoff that determines the ideal size of the diaphragm. If it is too
wide, the pressure drop is too small to be measured accurately, and
reed oscillations likely to occur. If the diaphragm is too small, the
system-wide characteristic can become too steep, making part of the
($(\Delta p)_r$) range inaccessible. 

The ideal diaphragm cross-section is then found empirically, by trying
out several resistance values until one complete measurement can be
done without oscillations or sudden closings of the reed. The optimal
diaphragm diameter is sought using a medical flow regulator with
continuously adjustable cross-section as a replacement for the
diaphragm.

% Initial problems:
%   diaphragm choice (reed shuts / reed sings)
% \rem{$\Downarrow$ René says: too much details}

% In practice, the experiment is first performed with a device replacing
% the diaphragm, whose resistance can be adjusted continuously until the
% system behaves conveniently. One possibility for this replacement,
% which we used, is a medical flow regulator used in intravenous
% injections. Once an appropriate resistance is found, this device is
% calibrated using the method described below for the diaphragms, and an
% appropriate diaphragm is sought whose characteristics approach that of
% the replacement device using the data in figure \ref{fig:diaph}.

\paragraph{Finer control of the mouth pressure $p_m$}

% \rem{$\Downarrow$ René says: too much details}

% During the attempts to find an optimal diaphragm, it was found that
% sudden closures were correlated to sudden increases in the mouth
% pressure. A part of the problem is that the mouth pressure is not
% controlled directly. This control is done using a pressure reducer
% between the main compressed air source and the experimental apparatus.
% The pressure in the mouth thus depends both on the valve position and
% on the downstream resistance.

% By introducing a leak between the pressure reducer and the
% experimental apparatus, it is possible to have smaller pressure
% increases in the mouth while the reed is closing. This improves
% measurements in the decreasing region of the characteristic
% (\textbf{BC}) when the system-wide characteristic is not multi-valued.
% The addition of a leak does not influence the experiment, because it
% is placed upstream of the artificial mouth.

% \rem{$\Uparrow$ René says: too much details}

During the attempts to find an optimal diaphragm, it was found that
sudden closures were correlated to sudden increases in the mouth
pressure. A part of the problem is that the mouth pressure depends
both on the reducer setting [or configuration] and on the downstream
resistance. By introducing a leak upstream of the experimental
apparatus (thus not altering the experiment) , it is possible to
improve measurements in the decreasing region of the characteristic
(\textbf{BC}), at least when the system-wide characteristic is not
multi-valued (see section \ref{sec:diaphr-reduc-range}).

\paragraph{Increase the reed mass}

One other way to reduce the oscillations is thus to prevent the
appearance of instabilities, or to reduce their effects. An increase
in the reed damping would certainly be a good method to avoid
oscillations, because it cancels out the active role of the reed
(which can be seen as a negative damping) \cite{Deb04}. 

It is difficult to increase the damping of the reed without altering
its opening or stiffness properties. The simplest way found to prevent
reed oscillations was thus an increase in the reed mass.

\begin{figure}[htbp]
  \centering
  \includegraphics[width=.7\columnwidth]{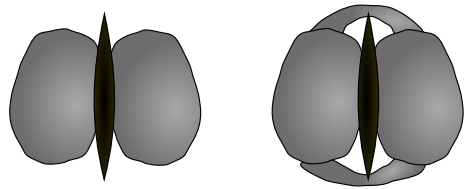}
%  \hspace{.1\columnwidth}
%  \includegraphics[width=.3\columnwidth]{oboe_reed_masses_devant_humide}
  \caption{Front view of the reed
    with attached masses, at left in dry conditions, at right in
    soaked conditions (to prevent the masses from slipping).}
  \label{fig:car_masses}
\end{figure}

This mass increase was implemented by attaching small masses of
\emph{Blu-Tack}\footnote{A plastic adherent material usually used to
  fixate paper to a wall} to one or both blades of the reed
(fig.~\ref{fig:car_masses}). During measurements on previously soaked
reeds it was difficult to keep the masses attached to the reed, so
that an additional portion of \emph{Blu-Tack} is used to connect the
to masses together, wrapping around the reed. A comparison of the
results using different masses showed that their effect on the
quasi-static characteristics can be neglected.

\subsection{Experimental set-up and calibrations}
\label{sec:car_proc}

\begin{figure}[htbp]
  \centering
%   \psfrag{pm}{$p_m$}
%   \psfrag{pr}{$p_r$}
  \includegraphics[width=1\columnwidth]{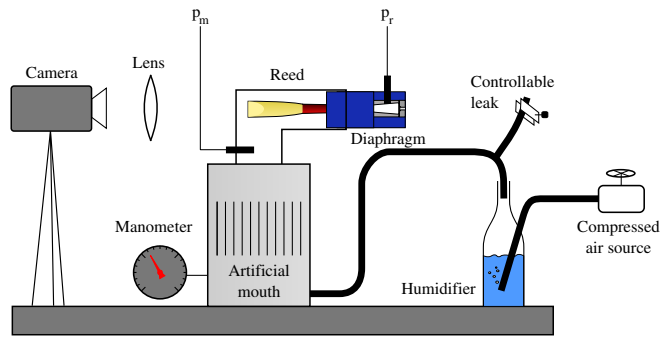}
  \caption{Device used for characteristics measurements.}
  \label{fig:car_schema}
\end{figure}

The experimental device is shown in figure \ref{fig:car_schema}. An
artificial mouth \cite{Alm04b} was used as a blowing mechanism and
support for the reed. The window in front of the reed allows the
capture of frontal pictures of the reed opening.

Artificial lips, allowing to adjust the initial opening area of the
reed were not used here, fearing that they would modify some of the
elastic properties of the reed, yet differently from what happens with
real lips.

As stated before, the plot of the characteristic curve requires two
coordinated measurements: the pressure difference $(\Delta p)_r$
across the reed and the induced volume flow $q$, determined from the
pressure drop $(\Delta p)_d$ across a calibrated diaphragm
(sect.~\ref{sec:car_flow_meas}).

% As seen
% above (sect.~\ref{sec:car_flow_meas}) the measurement of the volume
% flow was effectuated indirectly, by a measurement of a pressure drop
% in a calibrated diaphragm.

In practice thus, the experiment requires two pressure measurements
$p_m$ and $p_r$, as shown in figure \ref{fig:car_schema}.

% Eventually an additional measurement of the reed opening area was
% effectuated based on frontal video captures of the reed opening. 

% The experiments presented below consist in the measurement of two main
% quantities, pressure and flow 

% The most fundamental measurement we will aim is the characteristic
% curve which relates the pressure difference between the mouth and the
% inside of the reed (at the reed output) and the volume flow that
% crosses the reed.

% However, in order to have a better understanding about the details of
% the mechanic and aerodynamic details of the reed, flow and pressure
% measurements are often accompanied by a video recording of the reed
% opening from which will be extracted information about it, such as the
% area of the opening.

% Data about the reed opening is calculated using the same proceeding
% already described in section \ref{sec:im_anal}. 

% All experiments were performed in the artificial mouth described in
% appendix \ref{sec:artmouth}. No lips replacement was used.
% \paragraph{Image analysis}

% Reference to the chapter with the image analysis article

\subsubsection{Pressure measurements}

% Calibration of the pressure sensors
The pressure is measured in the mouth and in the reed using Honeywell
SCX series, silicon-membrane differential pressure sensors whose range
is from $-50$ to $50$ kPa.

These sensors are not mounted directly on the measurement points, but
one of the terminals in each sensor is connected to the measurement
point using a short flexible tube (about $20$ cm in length).
Therefore, one tube opens in the inside wall of the artificial mouth,
$4$ cm upstream from the reed, and the other tube crosses the rubber
socket attaching the diaphragm to the reed output.  The use of these
tubes does not influence the measured pressures as long as their
variations are slow.

The signal from these sensors is amplified before entering the digital
acquisition card. The gain is adjusted for each type of reed.  The
system consisting of the sensor connected to the amplifier is
calibrated as a whole in order to find the voltage at the amplifier
output corresponding to each pressure difference in the probe
terminals: the stable pressure drop applied to the probe is also
measured using a digital manometer connected to the same volumes, and
compared to the probe tension read using a digital voltmeter. Voltage
is found to vary linearly with the applied pressure within the
measuring range of the sensor.

% calibration of the lumped sensor plus amplifier system a simultaneous calibration of the total sensitivity of the sensor plus amplifier system is needed

\subsubsection{Diaphragm calibration}

\begin{figure}[htbp]
  \centering
  \includegraphics[width=.4\columnwidth]{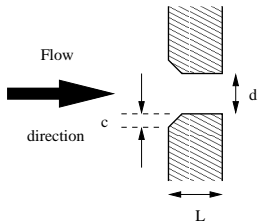}
  \caption
  {Detail of the diaphragm dimensions.}
  \label{fig:diaph_geom}
\end{figure}

The curve relating volume flow $q$ to the pressure difference through
diaphragms $(\Delta_p)_d$ can be approximated by the Bernoulli
theorem. In fact, diaphragms are constructed so as to minimize
friction effects (by reducing the length of the diaphragm channel) and
jet contraction --- the upstream edges are smoothed by chamfering at
45$^\circ$ (fig.~\ref{fig:diaph_geom}).  The chamfer height ($c$) is
approximately 0.5 mm. 
%This dimension is not very precise, because the chamfers are hand-made [Removed by René]
The diaphragm channel is 3 mm long ($L$).

\begin{figure}[htbp]
  \centering
  \includegraphics[width=.7\columnwidth]{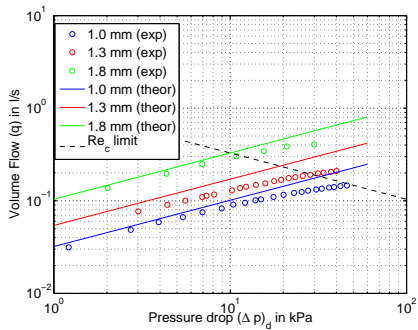}  
  \caption
  {Calibration of diaphragms used in characteristic measurements (dots
    are experimental data and lines are Bernoulli predictions using
    the measured diaphragm diameters). The dashed black line
    represents the pressure/flow relation corresponding to the
    expected transition between laminar and turbulent flows ($Re_c = 2000$),
    parameterised by the diaphragm diameter $d$.}
  \label{fig:diaph}
\end{figure}

Nevertheless, this ideal characteristic was checked for each diaphragm
(see fig~\ref{fig:diaph}). It was found that the effective
cross-section is slightly smaller than the actual cross-section (about
10\%), which is probably due to some \emph{Vena Contracta} effect in
the entrance of the diaphragm. Moreover, above a given pressure drop
the volume flow increase is lower than what is predicted by
Bernoulli's theorem (corresponding to a lower exponent than $1/2$
predicted by Bernoulli). This difference is probably due to turbulence
generated for high Reynolds Numbers. In figure \ref{fig:diaph} the
dashed line corresponding to the critical value of the Reynolds number
($Re_c=\frac{ud}{\nu}=2000$) is shown. It is calculated using the
following formulas for $u$ (the average flow velocity in the
diaphragm) and $d$ (the diaphragm diameter):
\begin{eqnarray}
  \label{eq:rec_d}
  d & = & \left(\frac{2}{\pi}\frac{q}{u}\right)^{1/2}\\
  u & = & \left(\frac{2 (\Delta p)_d}{\rho}\right)^{1/2}
\end{eqnarray}
so that the constant Reynolds relation is given by:
\begin{equation}
  \label{eq:rec_rel}
  Q^{1/2} \Delta p^{1/4} = Re_c \nu \left(\frac{\pi}{2}\right)^{1/2} \left(\frac{\rho}{2}\right)^{1/4}
\end{equation}
where the right-hand side should be a constant based on the diaphragm
geometry.

Since a suitable model was not found for the data displayed in figure
\ref{fig:diaph}, we chose to interpolate the experimental calibrations
in order to find the flow corresponding to each pressure drop in the
diaphragm. Linear interpolation was used in the $(p,q^2)$ space.

\subsubsection{Typical run}

% Time variation curve (pressures / area and pressure)
% Two different kinds of measurements can be distinguished:

% \begin{description}
% \item[Simple measurement] Only the pressures $p_m$ and $p_r$ are
%   measured, so that in the end only the pressure / flow characteristic
%   is available.
% \item[Measurement including opening data] 

% \rem{TO DO: Replace text with that of simple measurement in thesis and
%   remove the are line in fig. \ref{fig:car_time_cplx}}

\begin{figure}[htbp]
  \centering
  \includegraphics[width=.9\columnwidth]{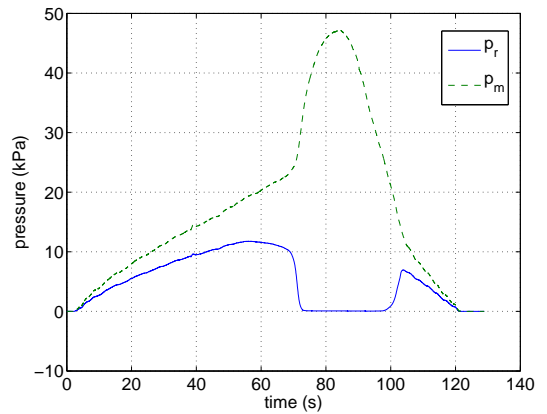}
%  \\\rem{To be replaced by a measurement without area?}
  \caption
  {Time variation of the mouth pressure ($p_m$)and the pressure inside
    the reed ($p_r$) during a successful characteristic measurement.}
  \label{fig:car_time_cplx}
\end{figure}

In a typical run, the mouth pressure $p_m$ is equilibrated with the
atmospheric pressure in the room in the beginning of the experiment.
Both $p_m$ and $p_r$ are recorded in the computer through a digital
acquisition device at a sampling rate of 4000 Hz. $p_m$ is increased
until slightly above the pressure at which the reed closes, left for
some seconds above this value and then decreased back to the
atmospheric pressure. The whole procedure lasts for about 3 minutes,
and is depicted in figure \ref{fig:car_time_cplx}.

% These pressure peaks are supposed to be simultaneous to the reed
% displacement peaks, because they are slow enough when compared to the
% natural period of oscillation of the reed (a fraction of a
% millisecond), and fast enough when compared to the visco-elastic
% relaxation times of the reed (the shortest being around 2 seconds
% \cite{Alm06b}).

% Once the images are analysed using the procedure described in a
% preceding article \cite{Alm06b}, the resulting geometrical data are
% oversampled (using linear interpolation) to the pressure data sample
% rate and the delay between both measurements is determined by hand,
% producing the red line in figure \ref{fig:car_time_cplx}.

\subsection{Double-reeds used in this study and operating conditions}
\label{sec:car_obj}

% Double-reeds come in a variety of forms, which was only roughly
% introduced in chapter \ref{sec:dblreed_inst}. For a single instrument
% such as the oboe, a great variety of materials or scrapping techniques
% can be found. It will be very interesting to 

Among the great variety of double-reeds that are used in music, we
chose as a first target for these measurements a natural cane
oboe-reed fabricated using standard procedures (by \emph{Glotin}), and
sold to the oboist (usually a beginner oboist) as a final product
(i.e., ready to be played).

%  \rem{Maybe there's a lot of people who
%   scrape the reed before using it, in order to tune it to their
%   preferences?}

% This choice was mainly based on considerations of what can be
% considered as an average reed, which can make a minimal consensus
% among all the different scraping techniques used by each instrumentist
% or reed-maker.

The choice of a ready-to-use cane reed was mainly retained because it
can be considered as an average reed. This avoids considering a
particular scraping technique among many used by musicians and
reed-makers. Of course, this does not greatly facilitate the task of
the reed measurement, because natural reeds are very sensitive to
environment conditions, age or time of usage.

Other reeds were also tested, as a term of comparison with the natural
reeds used in most of the experiments. However, none of these reeds
was produced by a professional oboist or reed maker, although it would
be an interesting project to investigate the variations in reeds
produced by different professionals.

To conclude, the results presented in the next section may depend to a
certain extent on the reed chosen for the experiments, and a larger
sample of reeds embracing the big diversity of scraping techniques
needs to be tested before claiming for the generality of the results
that will be presented.

Another remark has to be made on the conditions during the
experiments. The kind of reeds used in most experiments are always
blown with highly moisturised air.  In fact, in real life, reeds are
often soaked before they are used, and constantly maintained wet by
saliva and water vapour condensation.  These conditions were sought
throughout most of the experiments, although their sensitivity to
environmental conditions was also investigated. For instance, the
added masses were found to have no practical influence on the
nonlinear characteristics, whereas the humidity increases the
hysteresis in the complete measurement cycle (increasing followed by
decreasing pressures), while reducing the reed opening at rest
\cite{Alm06}.

%  \rem{ TO DO: The
%   dependence on humidity is not shown here, but maybe we could say a
%   word about it?  Talk about repeatability, influence of humidity and
%   added mass}

In our measurements, humidification is achieved by letting the air
flow through a plastic bottle half-filled with hot water at $40^\circ$
(see fig.~\ref{fig:car_schema}), recovering it from the top. Air
arriving in the artificial mouth has a lower temperature, because its
temperature is approximately $10^\circ$ when entering the bottle. This
causes the temperature and humidity to decrease gradually along the
experiments. Future measurements should include a thermostat for the
water temperature in order to ensure stable humidification

%   \rem{Drawbacks:
%     temperature decreases during experiment, so does humidification}

% All measurements were performed without artificial lips. The presence
% of lip force over the reed provides an offset of the pressure applied
% to the reed, which shows out in a reduction of the initial reed
% opening area $S_0$.

\section{Results and discussion}

\subsection{Typical pressure vs flow characteristics}
\label{sec:car_pq}

\begin{figure}[htbp]
  \centering
  \includegraphics[width=.9\columnwidth]{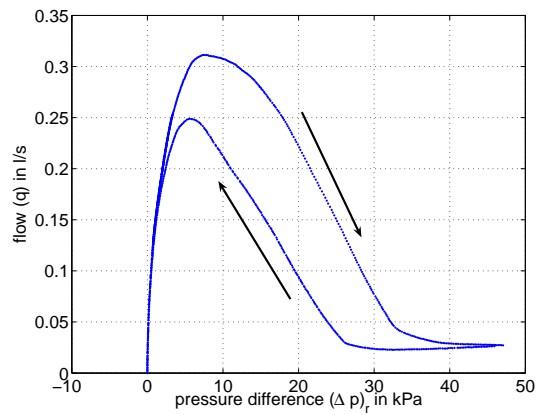}\\
  \caption
  {A typical result for the measurement of the volume flow vs pressure
    characteristic of a natural cane oboe reed.}
  \label{fig:car-typ}
\end{figure}

Using the formula of eq.~(\ref{eq:diaph}), and the calibrations
carried out for the diaphragm used in the measurement, the flow is
determined from the pressure inside the reed ($p_r$). The pressure
drop in the reed corresponds to the difference between the mouth and
reed pressures ($(\Delta p)_r = p_m-p_r$).  Volume flow ($q$) is then
plotted against the pressure difference ($(\Delta p)_r$), yielding a
curve shown in figure \ref{fig:car-typ}.

% A typical measurement (13, 14?)

% Figure \ref{fig:car-typ} is obtained for a natural cane oboe reed
% (from \emph{glottin}, used as bought without further scrapping). The
% air supplied to the instrument is humidified to 100\% so that the reed
% is tested in conditions that approach the ones used in real playing
% conditions.

In this figure, the flow is seen to increase until a certain maximum
value (at about 6 kPa). When the pressure is increased further, flow
decreases due to the closing of the reed. Instead of completely
vanishing for $(\Delta p)_r = p_M$, as predicted by the elementary
model shown in section \ref{sec:backuss-model}, the flow first
stabilises at a certain minimum value and then increases when the
pressure is increased further., indicating that it is very hard to
completely close the reed.

% Obs: vol flow stabilises before reaching q=0 as predicted by the
% elementary model, although the reed seems closed both through direct
% observations or image analysis. 

% The
% volume flow stabilises once a maximum pressure is attained
% (corresponding to the reed beating pressure $p_M$ in models) rather
% than being completely blocked by a closed reed.

% Observations of the reed opening made at the same time show that the
% two blades are in contact with each other at this point. It is
% possible that air is flowing through pores or irregularities in the
% blade's surfaces, which would require much higher pressures to be
% closed.

The flow remaining after the two blades are in contact suggests that
despite the closed apparency of the double reed, some narrow channels
remaining between the two blades are impossible to close, behaving
like rigid capillary ducts, which is corroborated by the slight
increase in the residual volume flow for high pressures. 

\begin{figure}[htbp]
  \centering
  \includegraphics[width=.9\columnwidth]{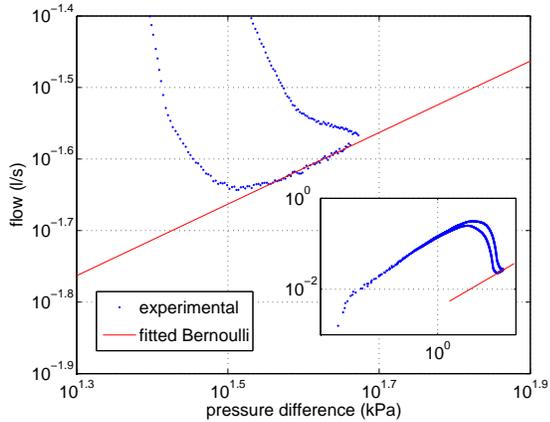}\\
  \caption
  {Logarithmic plot of the characteristic curve of figure
    \ref{fig:car-typ} to show the $1/2$ power dependence when the reed
    is almost shut.}
  \label{fig:car-log}
\end{figure}

If the residual reed opening is distributed over several channels, the
flow in each of these channels would be controlled by viscosity rather
then inertia, so that the volume flow $q$ should be proportional to
$(\Delta p)_r$ rather than to the $1/2$ power of it as in the case of
a Bernoulli flow. Nevertheless, observing a logarithmic plot of the
non-linear characteristics (fig.~\ref{fig:car-log}) shows a $1/2$
power dependence of the residual flow. This may mean that the reed
material is not yet stable at this point. 

% The fact that the flow does not decrease when the pressure is
% increased after the reed shutting suggests that this remaining opening
% stays more or less constant independently of the force applied to the
% reed. In fact, the flow tends to increase slightly, as expected when
% the opening remains constant and the pressure difference is increased.
% \rem{Note that the flow increase should be smaller than the one
%   predicted by the Bernoulli theorem, because these remaining openings
%   are probably very narrow and viscosity is probably very important.\\
%   TO DO: Verify the steepness of the curve above this point --- log/log plot.}

% \rem{Données de Guénel suggèrent que l'ouverture latérale pourrait
%   être plus importante quand l'anche est fermée, parce que la courbure
%   de l'anche est inversée à ce moment: chercher dans son rapport}

When reducing the pressure back to zero, the reed follows a different
path in the pressure/flow space than the path for increasing
pressures. This hysteresis is due to memory effects of the reed
material which have been investigated experimentally for single-reed
\cite{Dal03} and double-reed instruments \cite{Alm06b}.

\subsection{Comparison with other instruments}
\label{sec:comp-with-other}

\subsubsection{Bassoon}

Since oboes are not the only double-reed instruments, it is
interesting to compare the non-linear characteristic curves from
different instruments. The bassoon is also played using a double reed,
but its dimensions are different: its opening area at rest is
typically 7 mm$^2$ (against around 2 mm$^2$ for the oboe) and the
cross section profile varies slightly from oboe reeds.

% The bassoon is a typical case of another modern
% instrument that uses a different kind of double reed based on the same
% principle but with dimensions that are very different from those of
% the oboe. 

% \rem{Differences in reed dimensions from oboe to bassoon: opening area
%   (based on measurements done by image analysis) and variations in the
%   cross section profiles}

\begin{figure}[htbp]
  \centering
  \includegraphics[width=.9\columnwidth]{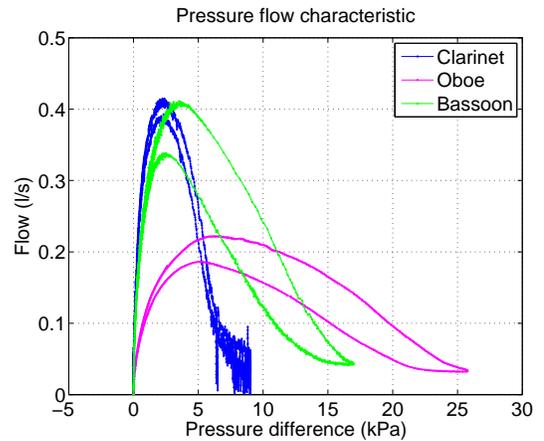}\\
  \rem{B\&W legend: increase dot size}
  \caption
  {Comparison of the characteristic curves of different reed exciters for different instruments. Clarinet data was obtained by S. Ollivier and J.-P. Dalmont \cite{Dal03} for a PlastiCover\textregistered\  reed. Oboe and bassoon reeds used moisturised air.}
  \label{fig:car-bsn}
\end{figure}

Figure \ref{fig:car-bsn} compares two characteristic curves for
natural cane oboe and bassoon reeds. Both were measured under similar
experimental conditions, as far as possible. The reed was introduced
dry in the artificial mouth, but the supplied air is moisturized at
nearly 100\% humidity, and masses were added to both reeds to prevent
auto-oscillations. The diaphragms used in each measurement are
different however, and this is because the opening area of the reed at
rest is much larger in the case of the bassoon, so that a smaller
resistance (a larger diaphragm) is needed to avoid that the reed
closes suddenly in the decreasing side of the characteristic curve
(see section \ref{sec:car_proc} and eq.~\ref{eq:15}). This should
not have any consequences in the measured characteristic curve.

% compare between Bh8 4h14 and clarinet dalmont

%The two experimental curves show similar behavior. 

In the $q$ axis, the bassoon reed reaches higher values, and this is
probably a consequence of its larger opening area at rest, although
the surface stiffness is likely to change as well from the oboe to
the bassoon reed. In the $p$ axis, the bassoon reed extends over
a smaller range of pressures so that the reed beating pressure is
about 17 kPa in the case of the bassoon reed whereas it is near 33 kPa
for the oboe reed.

\begin{figure}[htbp]
  \centering
  \includegraphics[width=.9\columnwidth]{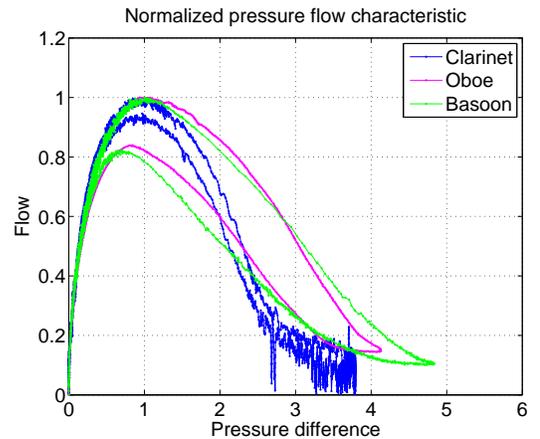}
%  \rem{B\&W legend: increase dot size}
  \caption
  {Normalized data from figure \ref{fig:car-bsn}}
  \label{fig:car-bsn-ad}
\end{figure}

Apart from these scaling considerations, the shape of the curves are
similar and this can be better observed if flow and pressure are
normalized using the maximum flow point of each curve (figure
\ref{fig:car-bsn-ad}).

% compare Bh8 4h14 and clarinet dalmont (normalized)

\subsubsection{Clarinet}
\label{sec:caract_snglreed}

The excitation mechanism of clarinets and saxophones share the same
principle of functioning with double reeds. However, there are
several geometric and mechanical differences between single reeds and
double reeds. For instance, flow in a clarinet mouthpiece encounters an
abrupt expansion after the first 2 or 3 millimeters of the
channel between the reed and the rigid mouthpiece, and the single reed
is subject to fewer mechanical constraints than any double reed. These
differences suggest that the characteristic curve of single-reed
instruments might present some qualitative differences with respect to
the double reed \cite{Ver03}. 

The non-linear characteristic curve of clarinet mouthpieces displayed
in figures \ref{fig:car-bsn} and \ref{fig:car-bsn-ad} was measured by
S. Ollivier an J.-P. Dalmont \cite{Dal03} using similar methods as the
ones we used for the double reed. A comparison between the curves for
both kinds of exciters (in figure \ref{fig:car-bsn}) shows that the
overall behavior of the excitation mechanism is similar in both cases.
Similarly to when comparing oboe to bassoon reeds, the scalings of the
characteristic curves of single-reeds are different from those of oboe
reeds, although closer to those of the bassoon.  This is probably a
question of the dimensions of the opening area.

A different issue is the relation between reference pressure values in
the curve (shown in the adimensionalized representation of figure
\ref{fig:car-bsn-ad}). As predicted by the elementary model described
in section \ref{sec:backuss-model}, in the single reed the pressure at
maximum flow is about $1/3$ of the beating pressure of the reed,
whereas in double-reed measurements, the relation seems to be closer
to $1/4$. This deviation from the model is shown in section
\ref{sec:analysis} to be linked with the diffuser effect of the
conical staple in double-reeds.

Figure \ref{fig:car-bsn-ad} also shows that in the clarinet mouthpiece
used by S.~Ollivier and J.~P.~Dalmont the hysteresis is relatively
less important than in both kinds of double-reeds. In fact, whereas
the measurements for double-reeds were performed in wet conditions,
the PlastiCover\textregistered\ reed used for the clarinet was
especially chosen because of its smaller sensitivity to environment
conditions. 

% \rem{What I would like to say here is something like (but this is a detail and would require that some graphics of synthetic reeds are included):\\
%   Plasticover\textregistered reeds are natural cane reeds with a
%   synthetic coating. The synthetic material is used to reduce the
%   gradual deformation of the reed due to the viscoelastic properties
%   of the cane. Nevertheless, if we compare the plasticover reed to a
%   fully synthetic oboe reed, the hysteresis seems stronger in the
%   latter case, which suggests that some of the hysteresis depends on
%   the geometry of the reed.}

% \rem{Even so, when comparing the PlastiCover\textregistered\ clarinet
%   reed to the synthetic double reed in figure \ref{fig:car-plast}, the
%   hysteresis seems more important for the double-reed. One of the
%   characteristics of the synthetic double-reed is also its greater
%   invariance with environment conditions. This would mean that there
%   might be some dependence on the geometry of the double reed (maybe
%   the pre-constraint) that affects the amplitude of the hysteresis
%   other than the intrinsic visco-elastic characteristics of the
%   material.}

\section{Analysis}
\label{sec:analysis}

\subsection{Comparison with the elementary model}
\label{sec:comp-with-back}

% \rem{Plot the characteristic curve along with a curve for Backus's model\\
%   Point the differences between both curves and suggest the pressure
%   recovery as an explanation for the shifting of the maximum-flow
%   pressure value}

The measured non-linear characteristic curve of figure
\ref{fig:car-typ} can be compared to the model described in section
\ref{sec:backuss-model}. In this model, two parameters ($k_s$ and
$S_0$) control the scaling of the curve along the $p$ and $q$ axis.
They are used to adjust two key points in the theoretical curve to the
experimental one: the reed beating pressure $p_M$ and the maximum
volume flow $q_\mathrm{max}$.

Once $q_\mathrm{max}$ is determined through a direct reading, the
stiffness $k_s$ is calculated using the following relation:
\begin{equation}
  \label{eq:1}
  k_s = q_\mathrm{max}^{-1} \left(\frac{2}{3}p_M\right)^{3/2} \rho^{-1/2}
\end{equation}

\begin{figure}[htbp]
  \centering \includegraphics[width=.9\columnwidth]{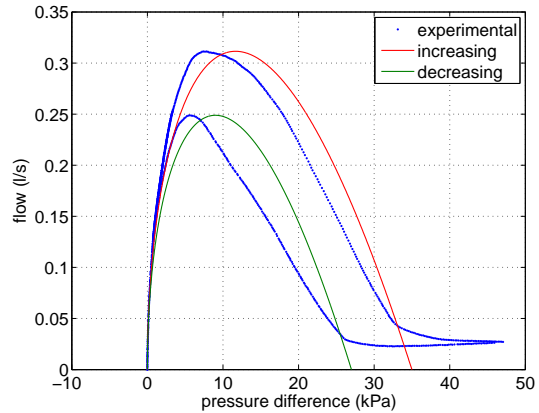}
%  \\\rem{Needs to be redone for B\&W, replace theoretical curves with
%    dotted/dashed curves.}
  \caption
  {Comparison of the experimental non-linear characteristics curve
    with the elementary model shown in figure \ref{fig:car_sr}. Two
    models are fitted, for increasing ($p_M=35$ kPa,
    $k_s=1.04\times10^{10}$ kg m$^{-3}$ s$^{-2}$) and decreasing ($p_M=27$ kPa,
    $k_s=8.86\times10^{9}$ kg m$^{-3}$ s$^{-2}$) mouth
    pressures.}
  \label{fig:car-backus}
\end{figure}

This allows to adjust a theoretical characteristic curve
(corresponding to the elementary model of equation
(\ref{eq:elem-model})) to each of the branches of the measured
characteristic curve, for increasing and decreasing pressures
(fig.~\ref{fig:car-backus}).

When compared to the elementary model of section
\ref{sec:backuss-model}, the characteristic curve associated to
double-reeds shows a deviation of the pressure at which the flow
reaches its maximum value. In fact, it can be easily shown that for
the elementary model this value is $1/3$ of the reed beating pressure
$p_M$, which is also verified in the clarinet (sect.
\ref{sec:caract_snglreed}). In the measured curves however, this value
is usually situated between $1/4 p_M$ and $1/5 p_M$.

Nevertheless, the shapes of the curves are qualitatively similar to
the theoretical ones. 

\subsection{Conical diffuser}
\label{sec:conical-diffuser}

\begin{figure}[htbp]
  \centering
  \hspace{-.05cm}
  \includegraphics[angle=0,width=.818\columnwidth]{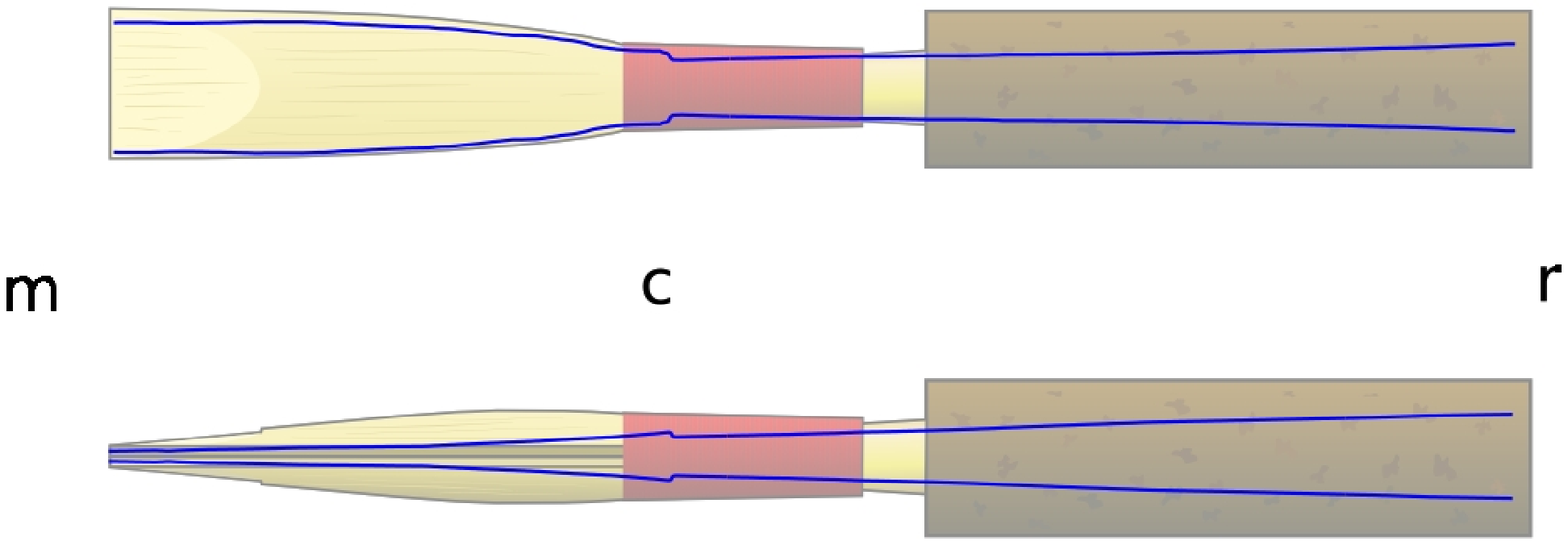}\\   
  \includegraphics[angle=0,width=.96\columnwidth]{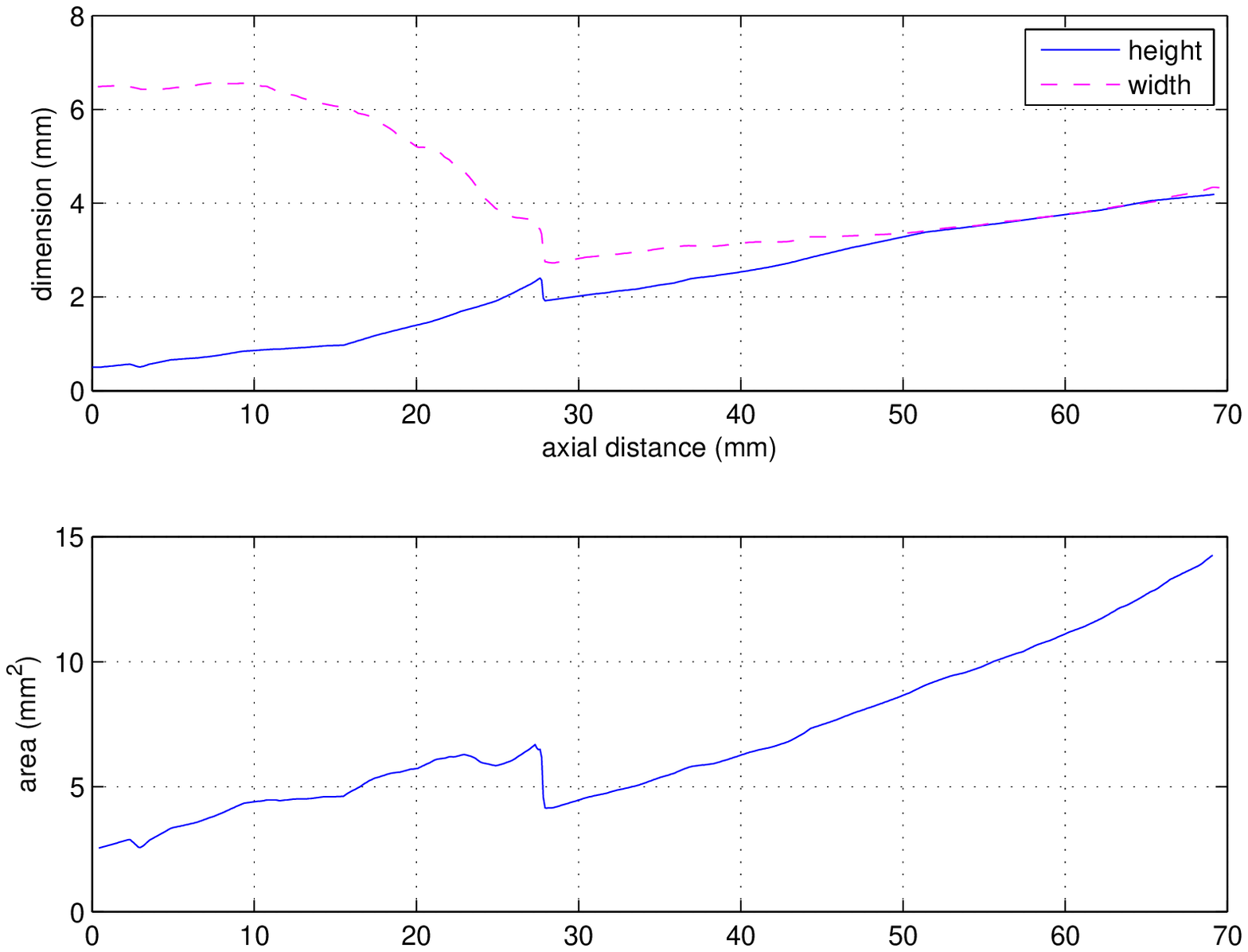}  
%\\  \rem{Index $c$ on figure}
  \caption{Cross-section profiles (axis
    and area) of an oboe reed, measured on a mold of the reed channel,
    and indexes used in section \ref{sec:conical-diffuser}:
    \textbf{m}outh, \textbf{c}onstriction and \textbf{r}eed output.}
  \label{fig:geom_eh}
\end{figure}

The former observations about the displacement of the maximum value
can be analysed in terms of the pressure recoveries due to flow
decelerations inside the reed duct. Variations in the flow velocity
are induced by the increasing cross-section of the reed towards the
reed output (fig.~\ref{fig:geom_eh}). This can be understood simply by
considering energy and mass conservation between two different sections
of the reed:
\begin{equation}
  \label{eq:3}
  p_{in} + \frac{1}{2} \rho \left(\frac{q}{S_{in}}\right)^2 = p_{out} + \frac{1}{2} \rho \left(\frac{q}{S_{out}}\right)^2 
\end{equation}
where $q$ is the total volume flow that can be calculated either at
the input or the output of the conical diffuser by integrating the
flow velocity over the cross-section $S_{in}$ or $S_{out}$
respectively.

In practice however, energy is not expected to be completely conserved
along the flow because of its turbulent nature. In fact, for instance
at the reed output (diameter $d$), the Reynolds number of the flow ($Re = \frac{u d}{\nu} = 4 \frac{q}{\pi d \nu}$) can be estimated using data from
figure \ref{fig:car-typ} to reach a maximum value of 5000. Given that
this number is inversely proportional to the diameter of the duct $d$,
the Reynolds increases upstream, inside the reed duct, so that the
flow is expected to be turbulent also for lower volume flows.

% In
% engineering literature, this duct geometry is known as a ``conical
% diffuser''

For turbulent flows, no theoretical model can be applied to calculate
the pressure recovery due to the tapering of the reed duct. However, 
phenomenological models are available in engineering literature, where
similar duct geometries are known as ``conical diffusers''. Unlike in
clarinet mouthpieces, where the sudden expansion of the profile is 
likely to cause a turbulent mixing without pressure recovery
\cite{Hir95}, this effect must be considered in conical diffusers. 
The pressure recovery is usually quantified in terms of a
\emph{recovery coefficient} $C_P$ stating the relation between the
pressure difference between both ends of the diffuser and the ideal
pressure recovery which would be achieved if the flow was stopped
without losses:
\begin{equation}
  \label{eq:flow_recov_def}
  C_P=\frac{p_\mathrm{out} - p_\mathrm{in}}{\frac{1}{2} \rho u_\mathrm{in}^2}
\end{equation}
$C_P$ values range from $0$ (no recovery) to $1$ (complete recovery,
never achieved in practice). Distributed losses due to laminar
viscosity along the reed are neglected.

According to equation (\ref{eq:flow_recov_def}), pressure recovery is
proportional to the square of the flow velocity at the entrance of the
conical diffuser, and consequently to the squared volume flow inside
the reed.  This is coherent with the shifting of the double-reed
non-linear characteristics curve for high volume flows, as observed in
figure \ref{fig:car-backus}.

% \rem{CHRISTOPHE proposes to add:\\
%   In fact, considering pressure recovery in the staple means that the
%   pressure difference $(\Delta p)_r$ between the mouth and the end of
%   the staple is smaller than the pressure difference between the mouth
%   and the staple entrance. This difference is all the more significant
%   as the velocity $u_{in}$, i.e., the volume flow $q=u_{in}S_{in}$ is
%   increasing.\\ I think this is not necessary...}

\rem{Try to make the above paragraph more clear}

\subsubsection{Reed model with pressure recovery}
\label{sec:backuss-model-recup}

In order to take into account the pressure recovery before the reed
output, the flow is divided into two sections, the upstream, until the
constriction at 28 mm (index $c$ in fig.~\ref{fig:geom_eh}) and the
conical diffuser part from the constriction until the reed output. In
the upstream section, no pressure recovery is considered, so that the
flow velocity can be calculated using the pressure difference between
the mouth and this point using a Bernoulli model, as in equation
(\ref{eq:Flow}), but replacing $p_r$ with $p_c$:
\begin{equation}
  \label{eq:Flow_rec}
   q = S \sqrt{\frac{2(p_m - p_c)}{\rho}} 
\end{equation}

Similarly, the reed opening is calculated using the same pressure
difference:
\begin{equation}
  \label{eq:spring_rec}
  (\Delta p)_c = p_m - p_c = k_S (S_0 - S)
\end{equation}

The total pressure difference used to plot the characteristic curve,
however, is different, because the recovered pressure has to be added
to $(\Delta p)_c$:
\begin{equation}
  \label{eq:4}
  p_m - p_r = (\Delta p)_c - C_p \frac{1}{2} \rho \left(\frac{q}{S_c}\right)^2
\end{equation}
where $S_c$ is the reed duct cross-section at the diffuser input
i.~e., at the constriction, which is found from figure
\ref{fig:geom_eh}, $S_c=4\times 10^{-6}$ m$^2$. 

\begin{figure}[htbp]
  \centering
  \includegraphics[width=.9\columnwidth]{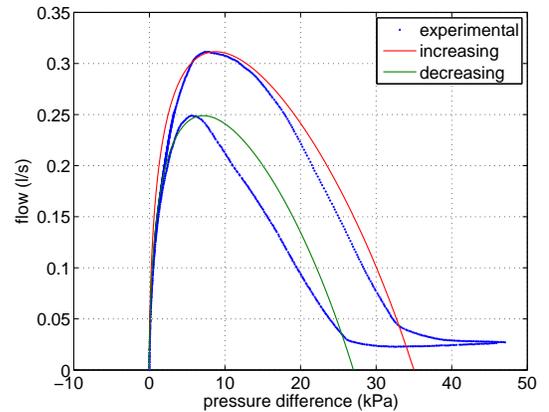}\\
  \caption
  {Comparison of the experimental non-linear characteristics curve
    with a reed model with pressure recovery in the final part of the
    duct. Fitted $k_s$ and $S_0$ are the same as in figure
    \ref{fig:car-backus} and different for increasing and decreasing
    mouth pressures. In both cases the value $C_p = 0.8$ was used.}
  \label{fig:car-recup}
\end{figure}

Using these equations, the modified model can be fitted to the
experimental data. In figure \ref{fig:car-recup}, the same parameters
$k_s$ and $S_0$ were used as in figure \ref{fig:car-backus}, leaving
only $C_p$ as a free parameter for the fitting. Figure
\ref{fig:car-recup} was obtained for a value of $C_p=0.8$.

This value can be compared to typical values of pressure recovery
coefficients found in industrial machines, for example in
\cite{Aza96}. 

%\paragraph{Experimental data found in engineering literature for the diffuser}

In engineering literature, $C_P$ is found to depend mostly on the
ratio between output and input cross-sections
($AR=S_\mathrm{out}/S_\mathrm{in}$) and the diffuser length to initial
diameter ratio ($L/d_\mathrm{in}$) \cite{Whi01}. The tapering angle
$\theta$ influences the growth of the boundary layers, so that above a
critical angle ($\theta=8^\circ$) the flow is known to detach from the
diffuser walls considerably lowering the recovered pressure.  An
in-depth study of turbulent flow in conical diffusers can be found in
the literature \cite{Aza96} usually for diffusers with much larger
dimensions than the ones found in the double reed.

The geometry of the conical diffuser studied in \cite{Aza96} can be
compared to the one studied in our work: the cross-section is circular
and the tapering angle $\theta=3.94^\circ$ is not very far from the
tapering angle of the reed staple $\theta=5.2^\circ$ (in particular,
both are situated in regions of similar flow regimes \cite{Kli62} as a
function of the already mentioned $AR$ and $L/d_\mathrm{in}$).
Reynolds numbers of his flows ($Re=6.9\times10^4$) are also close to
the maximum ones found at the staple input ($Re\simeq 10^4$).

The length to input diameter ratio of the reed staple $L/d=20$ is
bigger than that found in \cite{Aza96}, however the pressure recovery
coefficient can be extrapolated from his data to find the value
$C_P\simeq0.8$ (fig. 2 in \cite{Aza96}). \rem{(in \cite{Whi01},
  fig.~6.28b indicates $C_P=0.67$ for $L/d=20$ and $AR=4.5$)}

% \rem{Graphic demonstration of the reed cross-section profile.\\ 
% Role of the diffuser (pressure recovery, CITATIONS)\\ 
% Two options:\\
% \begin{itemize}
% \item Correct experimental data with the recovered pressure (same as in thesis)
% \item Fit a Backus Model + conical diffuser to the raw experimental data
% \end{itemize}
% }

\section{Conclusion}
\label{sec:conclusion}

The quasi-static non-linear characteristics was measured for double
reeds using a similar device as for single-reed mouthpieces
\cite{Dal03}. The obtained curves are close to the ones found for
single-reeds, and in particular no evidence of multivalued flows for a
same pressure was found, as was suggested by some theoretical
considerations \cite{Wij95}, \cite{Ver03}. 

Double-reed characteristic curves present substantial quantitative
differences for high volume flows when compared to elementary models
for the reed. These differences can be explained using a model of
pressure recovery in the conical staple, proportional to the input
flow velocity.

It is worth noting that direct application of this measurement to the
modelling of the complete oboe is not that obvious. Indeed, the
assumption that the mouthpiece as a whole (reed plus staple) can be
modelled as a non-linear element with characteristics given by the
above experiment would be valid if the size of the mouthpiece was
negligible with respect to a typical wavelength. Given the 7 cm of the
mouthpiece and the 50 cm \rem{???} of a typical wavelength, this is
questionable.  Should the staple be considered as part of the
resonator? In that case, the separation between the exciter and the
resonator would reveal a longuer resonator and an exciter without
pressure recovery.  This could be investigated through numerical
simulations by introducing the pressure recovery coefficient ($C_P$)
as a free parameter.

Moreover, the underlying assuption of the models is that non
stationnary effects are negligible (all flow models are quasi-static).
Some clues indicate that this could also be put into question:
\begin{itemize}
\item First of all, experimental observations of the flow at the
  output of the staple (through hot-wire measurements \cite{Alm06})
  revealed significant differences in the flow patterns when
  considering static and auto-oscillating reeds.
\item Moreover, non-dimensional analysis reveal Strouhal numbers much
  larger than for simple reed instruments (\cite{Ver03}, \cite{Alm06})
\end{itemize}

% At present it is not clear if the pressure recovery should be taken
% into account when modelling a complete instrument, because the
% resonator can be seen as a continuation of the reed staple. 

\begin{acknowledgments}
  The authors would like to thank J.-P. Dalmont for experimental data
  on the clarinet reed characteristics and fruitful discussions and
  suggestions about the experiments and analysis of data and A.
  Terrier and G. Bertrand for technical support.
\end{acknowledgments}

\appendix

\section{Calculation of the minimum diaphragm cross-section}
\label{sec:calc-minim-diaphr}

The total pressure drop in the reed-diaphragm system
(fig.~\ref{fig:diaph_sk}) is:

\begin{equation}
  \label{eq:6}
  (\Delta p)_s = (\Delta p)_r + (\Delta p)_d 
\end{equation}

\rem{Replace following paragraph with differentiaition of
  eq.~(\ref{eq:elem-model}) as proposed by Christophe. Check.}

The system's characteristics becomes multivalued when there is at
least one point on the curve where the slope is infinite:
\begin{equation}
  \label{eq:7}
  \frac{\partial}{\partial q}(\Delta p)_s = \frac{\partial}{\partial q}(\Delta p)_r + \frac{\partial}{\partial q}(\Delta p)_d = 0
\end{equation}

% The formula for $(\Delta p)_r$ (equation (\ref{eq:elem-model})) cannot
% be inverted analytically. 

Because of simplicity, the derivatives in equation (\ref{eq:6}) are
replaced by their inverse:
\begin{equation}
  \label{eq:8}
  \frac{\partial}{\partial q}(\Delta p)_s = \left(\frac{\partial q}{\partial (\Delta p)_r}\right)^{-1} + \left(\frac{\partial q}{\partial(\Delta p)_d}\right)^{-1}=0
\end{equation}
yielding
\begin{eqnarray}
  \label{eq:9}
  \left(\frac{S}{\rho}\left(\frac{2 (\Delta p)_r}{\rho}\right)^{-1/2} -
    \frac{1}{k_s}\left(\frac{2 (\Delta
        p)_r}{\rho}\right)^{1/2}\right)^{-1} + & & \nonumber\\
  + \frac{\rho}{S_d}\left(\frac{2 (\Delta p)_d}{\rho}\right)^{1/2} = 0
  & &
\end{eqnarray}

Solving for $S_d$:
\begin{eqnarray}
  \label{eq:10}
  S_d & =& - \left(\frac{2 (\Delta p)_d}{\rho}\right)^{1/2} \times
  \nonumber \\
  &\times& \left(\frac{S}{\rho}\left(\frac{2 (\Delta p)_r}{\rho}\right)^{-1/2} -
    \frac{1}{k_s}\left(\frac{2 (\Delta p)_r}{\rho}\right)^{1/2}\right)
\end{eqnarray}

Simplifying 
%and replacing $S=S_0-\frac{(\Delta p)_r}{k_s}$:
\begin{equation}
  \label{eq:11}
  S_d = -S \left(\frac{(\Delta p)_d}{(\Delta p)_r}\right)^{1/2} + \frac{2}{k_s}\left((\Delta p)_d (\Delta p)_r\right)^{1/2}
\end{equation}

From equations (\ref{eq:Flow}) and (\ref{eq:diaph}), we can find
\begin{equation}
  \label{eq:12}
  (\Delta p)_r = \left(\frac{S_d}{S}\right)^2 (\Delta p)_d
\end{equation}
and equation (\ref{eq:11}) can be written 
\begin{equation}
  \label{eq:13}
  S_d = -S\frac{S}{S_d} + \frac{2}{k_s}\frac{S}{S_d}(\Delta p)_r
\end{equation}

Now we can replace $S=S_0-\frac{(\Delta p)_r}{k_s}$ to find
\begin{equation}
  \label{eq:14}
  S_d^2 = (-S_0 + 3\frac{(\Delta p)_r}{k_s})(S_0 -\frac{(\Delta p)_r}{k_s})
\end{equation}

It is clear that the right-hand side of this equation must be
positive. Moreover, it is a parabolic function of $(\Delta S) =
\frac{(\Delta p)_r}{k_s}$, with its concavity facing downwards.

The maximum value of $S_d^2(S)$:
\begin{equation}
  \label{eq:15}
  \mathrm{max}\left(S_d^2(S)\right) = \frac{S_0^2}{3}
\end{equation}
is thus the value for which there is only a single point where the
characteristic curve has an infinite slope.

We thus conclude that $S_d = \frac{S_0}{\sqrt{3}} = 0.58 S_0$ is the
minimum value of the diaphragm cross-section that should be used for
flow measurements.

\bibliography{../../biblio/dblreed}

\begin{thebibliography}{}
\bibitem[end, b]{endnote2}
The flow velocity at the reed output is neglected when compared to the velocity
  inside the diaphragm ($S_{d} \ll S_{output}$).
\end{thebibliography}
\bibliographystyle{apalike}
%\bibliographystyle{unsrt}

%\newpage

%\listoffigures 

\end{document}